\documentclass{sig-alternate}

\usepackage{amsmath,amssymb,latexsym,cite}
\usepackage{pstricks,graphicx}

\newtheorem{definition}{Definition}

\newtheorem{corollary}{Corollary}
\newtheorem{lemma}{Lemma}
\newtheorem{theorem}{Theorem}

\newcommand{\beq}{\begin{equation}}
\newcommand{\eeq}{\end{equation}}
\newcommand{\bea}{\begin{eqnarray}}
\newcommand{\eea}{\end{eqnarray}}
\def\qed{\quad \vrule height6.5pt width6pt depth0pt} %Nifty end proof sign
\newcommand{\EE}{ \mathcal{E}} %Nifty Energy symbol

\newcommand{\diag}{\ensuremath{\operatorname{diag}}}

\long\def\symbolfootnote[#1]#2{\begingroup%
\def\thefootnote{\fnsymbol{footnote}}\footnote[#1]{#2}\endgroup}

\newcommand{\Wmat}{\ensuremath{W}}
\newcommand{\Pmat}{\ensuremath{P}}
\newcommand{\ones}{\ensuremath{\vec{1}}}
\newcommand{\defn}{\ensuremath{: \, = }}
\newcommand{\avec}{\ensuremath{\vec{a}}}

\newcommand{\qvec}{\ensuremath{\vec{q}}}
\newcommand{\yvec}{\ensuremath{\vec{y}}}
\newcommand{\eigtwo}{\ensuremath{\lambda_2}}

\newcommand{\order}{\ensuremath{\mathcal{O}}}
\newcommand{\rejeps}{\ensuremath{\mu}}
\newcommand{\rejdelta}{\ensuremath{\nu}}
\newcommand{\Prob}{\ensuremath{\mathbb{P}}}
\newcommand{\dur}{\ensuremath{\mathcal{D}}}

\newcommand{\widgraph}[2]{\includegraphics[keepaspectratio,width=#1]{#2}}
\newcommand{\Qnum}{\ensuremath{Q}}
\newcommand{\Exs}{\ensuremath{\mathbb{E}}}

\begin{document}
\conferenceinfo{IPSN'06,} {April 19--21, 2006, Nashville,
Tennessee, USA.}
\CopyrightYear{2006}
\crdata{1-59593-334-4/06/0004}

\title{Geographic Gossip:
Efficient Aggregation for Sensor Networks }
\numberofauthors{3}
\author{
%% e-mail address with \email.
\alignauthor Alexandros G. Dimakis\\
       \affaddr{Department of EECS} \\
       \affaddr{UC Berkeley}\\
       \email{\normalsize adim@eecs.berkeley.edu}
\alignauthor Anand D. Sarwate\\
 \affaddr{Department of EECS} \\
       \affaddr{UC Berkeley}\\
       \email{\normalsize asarwate@eecs.berkeley.edu}  
\alignauthor Martin J. Wainwright\\
 \affaddr{Department of EECS, and} \\
 \affaddr{Department of Statistics} \\
       \affaddr{UC Berkeley}\\
       \email{\normalsize wainwrig@eecs.berkeley.edu}
}

\maketitle

\begin{abstract}
Gossip algorithms for aggregation have recently received
significant attention for sensor network applications because of
their simplicity and robustness in noisy and uncertain
environments. However, gossip algorithms can waste significant
energy by essentially passing around redundant information
multiple times.  For realistic sensor network model topologies
like grids and random geometric graphs, the inefficiency of gossip
schemes is caused by slow mixing times of random walks on those
graphs.  We propose and analyze an alternative gossiping scheme
that exploits geographic information. By utilizing a simple
resampling method, we can demonstrate substantial gains over
previously proposed gossip protocols.  In particular, for random
geometric graphs, our algorithm computes the true average to
accuracy $1/n^a$ using $O(n^{1.5}\sqrt{\log n})$ radio
transmissions, which reduces the energy consumption by a
$\sqrt{\frac{n}{\log n}}$ factor over standard gossip algorithms.
\end{abstract}

%% A category with the (minimum) three required fields
%%A category including the fourth, optional field follows...
%\category{D.2.8}{Software Engineering}{Metrics}[complexity
%measures, performance measures]
%\terms{Delphi theory}
%\keywords{ACM proceedings, \LaTeX, text tagging}
%This saves more space

\vspace{1mm} \noindent {\bf Categories and Subject Descriptors:}
F.2.2, G.3

\vspace{1mm} \noindent {\bf General Terms:} algorithms

\vspace{1mm} \noindent {\bf Keywords:} gossip algorithms, random
geometric graphs, sensor networks, distributed consensus,
distributed aggregation

\section{Introduction}
Consider a network of $n$ sensors, in which each node collects a
measurement in some modality of interest (e.g., temperature,
light, humidity etc.).  It is frequently of interest to solve the
\emph{averaging problem}: namely, to develop a distributed and
fault-tolerant algorithm by which \emph{all nodes can compute the
average of all $n$ sensor measurements}.  Gossip algorithms solve
the averaging problem by having each node randomly pick one of
their one-hop neighbors and exchange their current values.  The
pair of nodes compute the pairwise average, which then becomes the
new value for both nodes.  By iterating this pairwise averaging
process, it is easy to show that all the nodes converge to the
global average in a completely distributed manner.  Although
fairly simple, the distributed averaging problem and related
consensus problems can be viewed as building blocks for solving
more complex problems~\cite{Kalman, Xiao}, including computing
general linear functions as well as optimization of non-linear
functions in sensor networks.

The key issue is how many iterations it takes for such gossip
algorithm to converge to a sufficiently accurate estimate.
Variations of this problem have received significant attention in
recent work \cite{Karp, Kempe, BoydInfocom, Chen}.  The
convergence speed of a nearest-neighbor gossip algorithm, known as
the \emph{averaging time}, turns out to be closely linked to the
mixing time of the Markov chain defined by a weighted random walk
on the graph. Boyd et al.~\cite{BoydInfocom} showed how to
optimize the neighbor selection probabilities for each node so to
find the fastest-mixing Markov chain on the graph. For certain
types of graphs, including complete graphs, expander graphs and
peer-to-peer networks, such Markov chains are rapidly mixing, so
that gossip algorithms converge very quickly.

Unfortunately, for the graphs corresponding to typical wireless
sensor networks, even an optimized gossip algorithm can result in
very high energy consumption.  For example, a common model for an
wireless sensor network is a random geometric
graph~\cite{Penrose}, in which all nodes communicate with
neighbors within a radius $r$.  With the transmission radius
scaling in the standard way as $r(n)=\Theta(\sqrt{\frac{\log
n}{n}})$, even an optimized gossip algorithm requires
$\Theta(n^2)$ transmissions (see section \ref{Comparisons}), which
is of the same order as the energy required for every node to
flood its value to all other nodes. This problem is noted
in~\cite{BoydInfocom}: ``In a wireless sensor network, Theorem 6
suggests that for a small radius of transmission, even the fastest
averaging algorithm converges slowly'', and it seems to be
fundamental for gossip algorithms on these graphs. Intuitively,
the nodes in a standard gossip protocol are essentially ``blind'',
and they repeatedly compute pairwise averages with their one-hop
neighbors. Information only diffuses slowly throughout the
network, roughly moving distance $\sqrt{k}$ in $k$ iterations (as
a random walk).

Accordingly, the goal of this paper is to develop and analyze
alternative ---and ultimately more efficient--- methods for
solving distributed averaging problems in wireless networks.  We
leverage the fact that \emph{sensors nodes typically know their
locations}, and can therefore use this knowledge to perform
geographic routing. Localization is a well studied problem
(e.g.,~\cite{Local1, Local2}), since geographic knowledge is
required in numerous applications.  With this perspective in mind,
we propose an algorithm that, like a standard gossiping protocol,
is completely randomized, distributed and robust, but requires
substantially less communication by exploiting geographic
information. The idea is that instead of exchanging information
with one-hop neighbors, geographic routing can be used to gossip
with random nodes who are far away in the network. We show that
the extra cost of multi-hop routing is compensated by the rapid
diffusion of information.

The remainder of this paper is organized as follows.  In Section
\ref{Algorithm}, we provide a precise statement of the distributed
averaging problem, describe our algorithm, and state our main
results on its performance. Section \ref{analysis} contains proofs
of these technical results.  In Section \ref{simulations}, we
experimentally evaluate the performance of our algorithm.

\section{Proposed Algorithm and Main Results}

\label{Algorithm}

\newcommand{\xave}{\ensuremath{\bar{x}_{\operatorname{ave}}}}

\subsection{Problem statement}
\label{SecProbState}

\subsubsection{Graph model} Following previous work~\cite{GK, BoydInfocom},
we model our wireless sensor network as a random geometric
graph~\cite{Penrose}. In this model, denoted $G(n,r)$, the $n$
sensor locations are chosen uniformly and independently in the
unit square, and each pair of nodes is connected if their
Euclidean distance is smaller than some transmission radius $r$.
(As discussed in Section~\ref{SecDiscussion}, our results have
natural analogs for lattices, and other graph structures that are
reasonable models of wireless networks).  It is well known
\cite{Penrose, GK, Gamal_tradeoff} that in order to have good
connectivity and minimize interference, the transmission radius
$r(n)$ has to scale like $\Theta ( \sqrt{\frac{\log n}{n}})$.  For
our analysis, we assume that communication within this
transmission radius always succeeds. Note however that the
proposed algorithm is very robust to communication and node
failures.

%Only include that if there is space
\begin{figure}
%\centering
%\includegraphics[width=9cm]{RGG_vor3.eps}
\widgraph{0.5\textwidth}{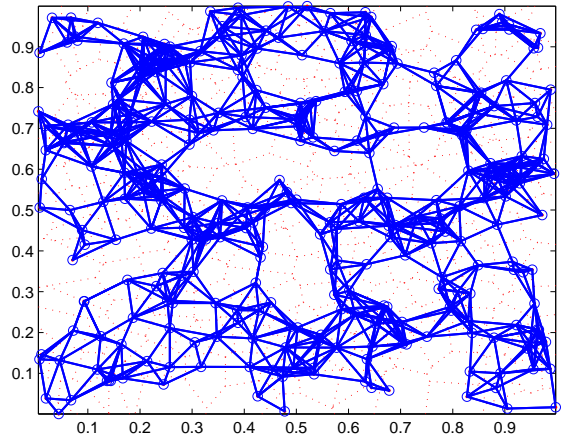} 
\caption{Illustration of a random geometric graph.  The solid lines
  represent graph connectivity, and the dotted lines show the Voronoi
  regions associated with each node.}
\end{figure}

\subsubsection{Time model} We use the asynchronous time
model~\cite{BoydInfocom}, which is well-matched to the distributed
nature of sensor networks.  More precisely, it is assumed that
each sensor node has a clock which ticks independently as a rate
$\lambda$ Poisson process.  Consequently, the inter-tick times are
exponentially distributed, and independent across nodes and across
time.  This set-up is equivalent to a single clock ticking
according to a rate $n\lambda$ Poisson process at times $Z_k$.  On
average, there are approximately $n$ clock ticks per unit of
absolute time (an exact analysis can be found in
\cite{BoydInfocom}) but we will always be measuring time in number
of ticks of this (virtual) global clock. Time is discretized, and
the interval $[Z_k, Z_{k+1})$ corresponds to the $k$th timeslot.
We can adjust time units relative to the communication time so
that only one packet exists in the network at each time slot with
high probability.

\subsubsection{Distributed averaging}
At time slot $k =0,1,2 \ldots$, each node $i=1, \ldots, n$ has an
estimate $x_i(k)$ of the global average, and we use $x(k)$ to
denote the $n$-vector of these estimates.  The ultimate goal is to
drive the estimate $x(k)$ to the average $\xave \ones$, where
$\xave \defn \frac{1}{n} \sum_{i=1}^n x_i(0)$, using the minimal
amount of communication. For the algorithms of interest to us, the
quantity $x(k)$ for $k > 0$ is a random vector, since the
algorithms are randomized in their behavior.
%%%%%%%%%%%%%%%%%%%%%%%%%%%%%%%%%%%%%%%%%%%%%%%%%%%%%%%%%%%%%%
Accordingly, we measure the convergence of $x(k)$ to $x(0)$ in the
following sense~\cite{Kempe,BoydInfocom} (essentially convergence
in probability):
\begin{definition}
Given $\epsilon > 0$, the $\epsilon$-averaging time is the
earliest time at which the vector $x(k)$ is $\epsilon$ close to
the normalized true average with probability greater than $1 -
\epsilon$: 
\beq \label{ave_time_definition}
T_{ave}(n,\epsilon)=\sup_{x(0)} \inf \left\{ k : \Prob \left(
\frac{ \|x(k)- x_{ave} \ones \|}{\|x(0)\|} \geq \epsilon \right)
\leq \epsilon \right\}~.
\eeq
\end{definition}
where $\| \cdot \|_2$ denotes the $\ell_2$ norm.

\newcommand{\CommCost}{\ensuremath{\mathcal{C}}}

Let $R(k)$ represent the number of one-hop radio transmissions
required for a given node to communicate with some other node at
time click $k$. In a standard gossip protocol, the quantity $R(k)
\equiv R$ is simply a constant, whereas for our protocol, $R(k)$
will be a random variable (with identical distribution for each
node).  The total communication cost is measured by the random
variable 
\beq 
\label{EqnDefnCommCost} \CommCost(n, \epsilon) = \sum_{k=1}^{T_{ave}(n,\epsilon)} R(k)~. 
\eeq 
In this paper, we
first analyze the expected communication cost, denoted by
$\EE(n,\epsilon)$, which is given by
\begin{align}
\EE(n,\epsilon) = E[R(k)] T_{ave}(n,\epsilon)~.
\end{align}
In addition, we provide a upper bound on the communication cost,
denoted by $\dur(n,\epsilon)$, such that 
\beq 
\Prob \Big \{ \CommCost(n, \epsilon) \geq  \dur(n, \epsilon) \Big \} \leq
\frac{\epsilon}{2}~. 
\eeq

\subsection{Proposed Algorithm}

The proposed algorithm combines gossip with geographic routing.
The key assumption is that each node knows its geographic
location.  With that knowledge, every node can also learn the
locations of its one-hop neighbors by having just one transmission
per node.

Suppose the $j$-th clock to tick belongs to node $s$.  Let $l(s)$
denote the location of node $s$.  Node $s$ activates and does the
following:
\begin{enumerate}
\item Node $s$ chooses a point uniformly in the unit square.  Call
this the target $t$.  Node $s$ forms the tuple $m_s = (x_s(j), l(s),
t)$.
\item \label{ggr} Node $s$ sends $m_s$ to its one-hop neighbor closest
to $t$, if any exists.  If node $r$ receives a packet $m_s$, it sends
$m_s$ to its one-hop neighbor closest $t$.  Greedy geographic routing
terminates when a node receives the packet and has no one-hop
neighbors with distance smaller to the random target that its own.
Let $v$ be the node closest to $t$.
\item \label{rej} Node $v$ makes an independent randomized decision
to accept $m_s$. If the packet is accepted, $v$ computes its new
value $x_v(j+1) = (x_v(j) + x_s(j))/2$ and a message $m_{v} =
(x_{v}(j), l(v), l(s))$ is sent back to $s$ via greedy geographic
routing.  Node $s$ computes $x_s(j+1) = (x_v(j) + x_s(j))/2$, and
the round ends.
\item If the packet is rejected, $v$ chooses a new point $t'$ uniformly
in the plane and repeats steps \ref{ggr}--\ref{rej} with message
$m'_s
 = (x_{s}(j), l(s), t')$.
\end{enumerate}

We will refer to this procedure as a \emph{gossip round}.  Our
analysis of this randomized algorithm, given in
Section~\ref{analysis}, consists of the following steps.  First,
we prove that when $r(n)= \Theta (\sqrt{\frac{\log n}{n}})$,
greedy routing always reaches the closest node $v$ to the random
target in $O(\sqrt{\frac{n}{\log n }})$ radio transmissions. Note
that in practice more sophisticated geographic routing algorithms
(e.g., \cite{GPSR}) can be used to ensure that the packet
approaches the random target when there are ``holes'' in the node
density. However, greedy geographic routing is good enough for our
model and other choices for routing algorithms will not affect our
results.

Our randomized procedure induces a probability distribution over
the chosen sensor $v$ (i.e., the one closest to the randomly
chosen target).  If this distribution were uniform, then it
follows immediately that the averaging time $T_{ave}(n, \epsilon)$
is $\order(n \log \epsilon^{-1})$.  In actuality, the probability
of choosing sensor $v$ is equal to $a_v$, the area of its
associated Voronoi region.  The distribution of Voronoi regions is
not very uniform, so in order to bound the averaging time
$T_{ave}(n, \epsilon)$, we apply rejection sampling  in order to
temper the distribution.  In particular, we apply the following
rejection sampling scheme, due to Bash et al.~\cite{rejsamp}.  Let
$\avec$ be an $n$-vector of areas of the sensors' Voronoi regions.
We set a threshold $\tau$ on the cell areas.  Sensors with cell
area smaller than $\tau$ always accept a query, and sensors with
cell areas larger than $\tau$ reject the query with a certain
probability.  The rejection sampling method protects against
oversampling and limits the number of undersampled sensors, and
allows us to prove that $T_{ave}(n, \epsilon) = \order(n \log
\epsilon^{-1})$, even for this perturbed distribution.

Of course, the rejection sampling scheme requires some random
number $\Qnum$ of queries before a sensor accepts.  In terms of
the number of queries, the total number of radio transmissions for
the $k$th gossip round is \beq \label{Round_duration} R(k)=
O\left(\Qnum \sqrt{\frac{n}{\log n}}\right). \eeq
Therefore if $T_{ave}$ gossip rounds take place overall, the
expected of radio transmissions will be 
\beq 
\EE(n, \epsilon) = \Exs [\Qnum] \order \left(\sqrt{\frac{n}{\log n}}\right)
T_{ave}(n,\epsilon)~. 
\eeq 
Accordingly, a third key component of
our analysis in Section~\ref{analysis} is to show that the
probability of acceptance remains \emph{larger than a constant},
which allows us to upper bound the expectation of the geometric
random variable $\Qnum$.  We also prove an upper bound on the
maximum value of $\Qnum$ over $T_{ave}$ rounds that holds with
probability greater than $1 - \epsilon/2$.

Putting these pieces of the analysis together, the main result of
this paper is that under the proposed geographic gossip algorithm
\beq T_{ave}(n,\epsilon)= O( n \log (1/\epsilon)) \eeq and
therefore the total cost for computing the average with geographic
gossip is \beq \label{EqnExpMain} \EE(n,\epsilon)= O
\left(\frac{n^{3/2}}{\sqrt{\log n}} \log \epsilon^{-1} \right).
\eeq Moreover, note that if we set $\epsilon = 1/n^\alpha$ in
equation~\eqref{EqnExpMain}, then we obtain $\EE(n,1 / n^\alpha
)=O \left(n^{3/2} \sqrt{\log n} \right)$.

\subsection{Related work and Comparisons}
\label{Comparisons}

In a series of papers \cite{BoydInfocom,BoydCDC}, Boyd et al. have
analyzed the performance of standard gossip algorithms.  Their fastest
standard gossip algorithm for the ensemble of random geometric graphs
$G(n,r)$ has a $\epsilon$-averaging
time\cite{BoydInfocom}\footnote{This quantity is computed in section
  IV.A of \cite{BoydInfocom} but the result is expressed in terms of
  absolute time units which needs to be multiplied by $n$ to become
  clock ticks.}  $T_{ave}(n, \epsilon)= \Theta ( n \frac{\log
  \epsilon^{-1} }{r(n)^2})$.  For the $r(n)$ in this paper this
averaging time is $\Theta (\frac{n^2}{\log n} \log \epsilon^{-1})$.
For $\epsilon$ scaling like $n^{-a}$ for any $a > 0$, this averaging
time scales likes $\Theta(n^2)$.  Note that in standard gossip, each
gossip round corresponds to communication with only one-hop neighbor
and hence costs only one radio transmission which means that the
fastest standard gossip algorithm will have a total cost $\EE(n)=
\Theta (n^2)$ radio transmissions for $\epsilon = \Theta(n^{-a})$.
Therefore, our proposed algorithm saves a factor of
$\sqrt{\frac{n}{\log n}}$ in communication energy by exploiting
geographic information.

Two very recent papers by Moallemi and Van Roy \cite{Moallemi} and
Mosk-Aoyama and Shah \cite{MoskAoyama} also consider the problem
of computing averages in networks.  The consensus propagation
algorithm of \cite{Moallemi} is a modified form of belief
propagation that attempts to mitigate the inefficiencies
introduced by the ``random walk'' in gossip algorithms.  However,
their results, although promising, have only been proven for
regular graphs, and it is unclear whether their algorithm will
prove efficient for the networks in this paper.  In
\cite{MoskAoyama}, the authors use an algorithm based on Flajolet
and Martin \cite{flajolet} to compute averages and bound the
averaging time in terms of a ``spreading time'' associated with
the communication graph.  However, they only show the optimality
of their algorithm for a graph consisting of a single cycle, so it
is currently difficult to speculate how it would perform on a
geometric random graph.

In \cite{Alanyali} the authors consider the related problem of
computing the average of a network in \emph{a single node}. They
propose a distributed algorithm to solve this problem and show how
it can be related to cover times of random walks on graphs.

\section{Analysis}

\label{analysis}

\subsection{Routing in $O(1/r(n))$}

We first need some simple lemmas about the network connectivity and
the feasibility of greedy geographic routing.

\begin{lemma}[Network connectivity]
\label{Connectivity} Let a graph be drawn randomly from the
geometric ensemble $G(n,r)$ defined in Section~\ref{SecProbState},
and a partition be made of the unit area into squares of length $\alpha
(n)= \sqrt{ 2 \frac{\log n }{n} }$.  Then the following statements
all hold with high probability:
\begin{enumerate}
\item[(a)] Each square contains at least one node.
\item[(b)] If $r(n)= \sqrt{ 10 \frac{\log n }{n} }$, then each
node will be able to communicate to a node in the four adjacent
squares.
\item[(c)] All the nodes in each square are connected with each other.
\end{enumerate}
\end{lemma}
\begin{proof}
The proof of part (a) following easily since it requires $\Theta
(n \log n)$ balls thrown randomly to cover $n$ bins with high
probability. (See \cite{MR95} and \cite{Gamal_tradeoff} for more
details). Moreover, if we select $r(n)= \sqrt{5} \alpha(n)$, then
simple geometric calculations show that each node will be able to
communicate to all other nodes in its square, as well as all nodes
in the four adjacent squares.
\end{proof}

\begin{lemma}[Greedy geographic routing]
  \label{routing} Suppose \\ that a node target location is chosen in the
  unit square.  Then greedy geographic routing will route to the node
  closest to the target in $O( 1/ r(n) )= O(\sqrt{ \frac{n}{\log n}})$
  steps.
\end{lemma}
\begin{proof}
By Lemma~\ref{Connectivity}(a), every square of of side length
$\alpha (n)= \sqrt{ 2 \frac{\log n }{n} } $ is occupied by at
least a node. Therefore, we can perform greedy geographic routing
by first matching the row and then the column of the square which
contains the target, which requires at most $\frac{2}{
r(n)}=O(\sqrt{ \frac{n}{\log n}})$ hops.  After reaching the
square where the target is contained, Lemma~\ref{Connectivity}(c)
guarantees that the subgraph contained in the square is completely
connected.  Therefore, one more hop suffices to reach the node
closest to the target.
\end{proof}

These routing results allow us to bound the cost in hops for an
arbitrary pair of nodes in the network to exchange values.  In the
next section, we describe a rejection sampling method used to
reduce the nonuniformity of the distribution (induced by sampling
locations rather than sensors).

%%%%%%%%%%%%%%%%%%%%%%%%%%
%%% REJECTION SAMPLING %%%
%%%%%%%%%%%%%%%%%%%%%%%%%%
\subsection{Rejection sampling}

As mentioned in the previous section, sampling geographic
locations uniformly induces a nonuniform sampling distribution on
the sensors in which a sensor $v$ is queried with probability
proportional to the area $a_v$ of its Voronoi cell.  However, by
judiciously rejecting queries, the sensors with larger Voronoi
areas can ensure that they are not oversampled.  We adopt the
following sampling scheme~\cite{rejsamp}: given some threshold
$\tau > 0$, sensor $v$ \textit{accepts} the request with
probability
\begin{eqnarray}
r_v & = & \min\left(\frac{\tau}{a_v}, 1\right)~.
\end{eqnarray}
\noindent We can then calculate the probability $q_v$ that sensor
$v$ is sampled:
\begin{eqnarray}
q_v & = & \frac{ \min(\tau, a_v) }{ \sum_{t=1}^{n} \min(\tau, a_t)
} \nonumber
\\ & = & \frac{ \min(\tau, a_v) }{ |\{t : a_t \ge \tau\}| \cdot \tau +
\sum_{t : a_t < \tau} a_t}~.
\end{eqnarray}
\noindent Of more importance to us is the denominator of $q_v$, which is the
total chance that a query is accepted:
\begin{align}
P_a = \sum_{v=1}^{n} a_v \min\left(\frac{\tau}{a_v}, 1\right) =
|\{v : a_v \ge \tau\}| \tau + \sum_{v : a_v < \tau} a_v~.
\end{align}
\noindent Let $\Qnum$ denote the total number of requests made by
a sensor before one is accepted.

%--------------------rejfig---------------------

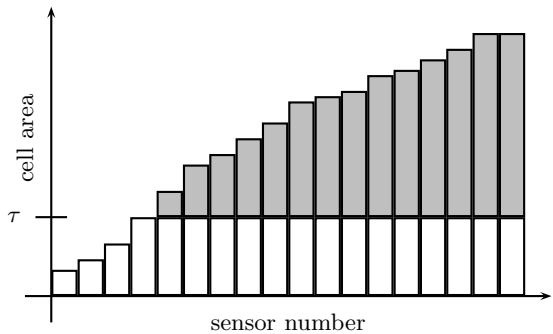
\begin{figure}
\begin{center}
\psset{xunit=0.07cm,yunit=0.07cm,runit=0.07cm}
\begin{pspicture}(-10,-10)(95,55)
\newgray{ltgray}{0.75}

\psline{->}(-5,0)(95,0) \psline{->}(0,-5)(0,55) \psframe(0,0)(5,5)
\psframe(5,0)(10,7) \psframe(10,0)(15,10) \psframe(15,0)(20,15)
\psframe(20,0)(25,15) \psframe(25,0)(30,15) \psframe(30,0)(35,15)
\psframe(35,0)(40,15) \psframe(40,0)(45,15) \psframe(45,0)(50,15)
\psframe(50,0)(55,15) \psframe(55,0)(60,15) \psframe(60,0)(65,15)
\psframe(65,0)(70,15) \psframe(70,0)(75,15) \psframe(75,0)(80,15)
\psframe(80,0)(85,15) \psframe(85,0)(90,15)

\psframe[fillstyle=solid,fillcolor=lightgray](20,15)(25,20)
\psframe[fillstyle=solid,fillcolor=lightgray](25,15)(30,25)
\psframe[fillstyle=solid,fillcolor=lightgray](30,15)(35,27)
\psframe[fillstyle=solid,fillcolor=lightgray](35,15)(40,30)
\psframe[fillstyle=solid,fillcolor=lightgray](40,15)(45,33)
\psframe[fillstyle=solid,fillcolor=lightgray](45,15)(50,37)
\psframe[fillstyle=solid,fillcolor=lightgray](50,15)(55,38)
\psframe[fillstyle=solid,fillcolor=lightgray](55,15)(60,39)
\psframe[fillstyle=solid,fillcolor=lightgray](60,15)(65,42)
\psframe[fillstyle=solid,fillcolor=lightgray](65,15)(70,43)
\psframe[fillstyle=solid,fillcolor=lightgray](70,15)(75,45)
\psframe[fillstyle=solid,fillcolor=lightgray](75,15)(80,47)
\psframe[fillstyle=solid,fillcolor=lightgray](80,15)(85,50)
\psframe[fillstyle=solid,fillcolor=lightgray](85,15)(90,50)

\rput(45,-5){sensor number} \rput{90}(-5,30){cell area}

\psline(-3,15)(3,15) \rput(-7,15){$\tau$}

\end{pspicture}
\caption{Rejection sampling in pictures.  The total shaded area is
the probability of a query being rejected.  The new sampling
distribution is given by the white histogram, appropriately
renormalized.}
\end{center}
\label{rejsampfig}
\end{figure}
%--------------------------------end of rejfig--------------------

A graphical picture of rejection sampling on the graph of Voronoi
cells is shown in Figure \ref{rejsampfig}.  Rejection sampling
``slices'' the histogram at $\tau$, and renormalizes the distribution
accordingly.  The total area that is sliced off is equal to $1 -
P_{a}$, the probability that a query is rejected.  Thus we can see
that if $\tau$ is chosen to be too small, the probability of rejection
will become very large.  In Lemma~\ref{LemGeo} we show that choosing
$\tau = \Theta(n^{-1})$ will keep the rejection probability suitably
bounded away from $1$, so that the expected number of queries
$\Exs[\Qnum]$ will be finite.  In particular, we choose $\tau$ such
that
\begin{align}
\Prob(a_v \le \tau) = \min\left(\rejdelta, \frac{\rejeps}{1 +
\rejeps}\right)~.
\end{align}
\noindent The constants $\rejdelta$ and $\rejeps$ control the
undersampling and oversampling respectively.  With this choice of
$\tau$, the results of Bash et al.~\cite{rejsamp} ensure that no
sensor is sampled with probability greater that $(1 + \rejeps)/n$
and no more than $\rejdelta n$ sensors are sampled with
probability less than $1/n$. The following result establishes that
the acceptance probability remains sufficiently large:

%-----------------geomfig------------------------------------------

\begin{figure}
\begin{center}
\psset{xunit=0.07cm,yunit=0.07cm,runit=0.07cm}
\begin{pspicture}(0,0)(60,60)
\newgray{ltgray}{0.75}

\pscircle*(30,30){1} \pscircle[linestyle=dotted](30,30){10}
\pscircle*(18,46){1}
\psline(36,47)(49,34)(37,16)(14,18)(12,29)(36,47)
\psline{->}(30,30)(18,46) \psline{->}(30,30)(30,20)
\rput(35,25){$r$} \pscircle[linestyle=dotted](30,30){20}

\pscircle*(22,4){1} \pscircle*(4,20){1} \pscircle*(55,18){1}
\pscircle*(55,55){1}

\psline(36,47)(36,60) \psline(49,34)(60,36) \psline(37,16)(50,0)
\psline(14,18)(4,0) \psline(12,29)(0,35)

\end{pspicture}
\caption{Inscribing circles in Voronoi cells.} \label{circlebound}
\end{center}
\end{figure}
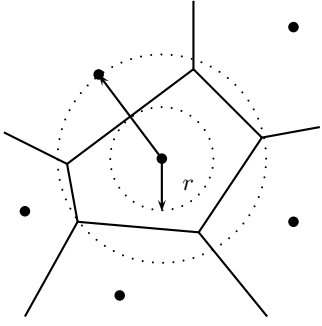
%-----------------------------------------------------------------

\begin{lemma}
\label{LemGeo} For $\tau = c n^{-1}$, we have $\Prob(a_v > \tau)
\geq 1 - 4 c$.
\end{lemma}

\begin{proof}
We use a simple geometric argument to lower bound $P(a_v > \tau)$.
Consider a node $s$ such that a circle of area $\tau$ it lies
entirely within its Voronoi region, as shown in
Figure~\ref{circlebound}.  Clearly, such nodes are a subset of
those with area larger than $\tau$.  Let the radius of this circle
be $r$ This $r$ is at most twice the distance to the closest node.
Thus in order to inscribe a circle of radius $\tau$ in the Voronoi
region, all other nodes must lie outside a circle of radius $2r$
around the node. This larger circle has area $4 \tau$, so
\begin{equation}
\Prob(a_v > \tau) \ge (1 - 4 \tau)^{n-1} = (1 - 4 c n^{-1})^{n-1}
\geq 1 - 4 c.
\end{equation}
\noindent
Thus, by appropriate choice of $c$, we can make the acceptance
probability arbitrarily close to $1$.
\end{proof}

Our next step is to bound the distance between the new sampling
distribution $\qvec$ and the uniform distribution $n^{-1}\ones$.
This will be used in next section to bound the second eigenvalue
of a matrix associated with the gossip algorithm.
\begin{lemma}
\label{rejnorm} For any $\epsilon > 0$, there exists constants
$\rejeps > 0 $ and $\rejdelta > 0$ such that rejection sampling
with parameters $(\rejeps, \rejdelta)$ leads to
\begin{subequations}
\begin{eqnarray}
\label{Eqnlonebound}
  \left\|\qvec - \frac{1}{n} \ones \right\|_1 & < & \epsilon \\
\label{Eqnltwobound}
  \left\|\qvec - \frac{1}{n} \ones \right\|_2 & < & \frac{1}{\sqrt{n}}
  \epsilon~.
\end{eqnarray}
\end{subequations}
\end{lemma}
\begin{proof}
Given $\epsilon > 0$, choose $\rejdelta$ and $\rejeps$ such that
$\rejdelta + \rejeps < \epsilon$ and $\rejdelta + \rejeps^2 <
\epsilon^2$. We then expand the error function and use the
properties given by the sampling scheme.
\begin{align*}
\sum_{v=1}^{n} \left| q_v - \frac{1}{n} \right| \le \sum_{v: a_v <
\tau} \left| q_v - \frac{1}{n} \right| + \sum_{v : a_v \ge \tau}
\left| q_v - \frac{1}{n} \right|
\end{align*}
\noindent Now we use the properties of rejection sampling.  On the
set $\{v : a_v < \tau\}$ we have $\frac{1}{n} > \tau$, so we can
upperbound the error by $\frac{1}{n}$.  Furthermore, we know $|\{v
: a_v < \tau\}| < \rejeps n$.  On the set $\{v : a_v \ge \tau\}$
we know $q_v$ is constant and $1/n \le q_v \le (1 + \rejdelta)/n$
by construction. Thus
\begin{align*}
\sum_{v=1}^{n} \left| q_v - \frac{1}{n} \right| &\le \left( \rejdelta
n \frac{1}{n} + n \left(\frac{1 + \rejeps}{n} - \frac{1}{n} \right)
\right) \\ &\le \rejdelta + \rejeps,
\end{align*}
which is less than $\epsilon$ by our choice of $\rejdelta$ and
$\rejeps$.

Turning now to the bound~\eqref{Eqnltwobound}, we write
\begin{align*}
\left\|\qvec - \frac{1}{n} \ones \right\|_2^2 &= \sum_{v: a_v < \tau} \left| q_v - \frac{1}{n} \right|^2 + \sum_{v : a_v \ge \tau} \left| q_v - \frac{1}{n} \right|^2 \\
&\le \rejdelta n \frac{1}{n^2} + n \left(\frac{\rejeps}{n}\right)^2 \\
&\le \frac{1}{n} (\rejdelta + \rejeps^2) \\ &\le \frac{1}{n}
\epsilon^2~.
\end{align*}
\end{proof}

Finally, we need to bound the expected number of rejections and the
maximum number of rejections in order to bound the expected number of
transmissions and total transmission time.  Recall that $\Qnum$ is the
number of queries that a sensor has to make before one is accepted,
and has distribution
\begin{equation}
\Prob(\Qnum = t) = P_a (1-P_a)^{t-1}.
\end{equation}

\begin{lemma}
\label{exp_rej} For a fixed $(\rejeps, \rejdelta)$, rejection
sampling leads to a constant number of expected rejections.
\end{lemma}
\begin{proof}
The random variable $\Qnum$ is just a geometric random variable
with parameter $P_{a}$, so we can write its mean as:
\begin{align*}
\Exs[\Qnum] &= \sum_{j=1}^{\infty} j (1 - P_a)^{j-1} P_a  \\
&= \frac{1}{P_a} \\ 
&= \frac{1}{|\{v : a_v \ge \tau\}| \tau + \sum_{v : a_v < \tau} a_v} \\ 
&\le \frac{1}{(1 - \rejdelta) \tau
n} = \order(1)~.
\end{align*}
since $\tau = \Theta(n^{-1})$ by construction.
\end{proof}

\begin{lemma}
\label{max_rej} Let $\{\Qnum_k : k = 1, 2, \ldots K\}$ be a set of
iid random variables identitically distributed according to $\Qnum$.
For a fixed $(\rejeps, \rejdelta)$, rejection sampling gives
\begin{equation}
\max_{1 \le k \le K} \Qnum_k = O(\log K + \log \epsilon^{-1})
\end{equation}
\noindent with probability greater than $1 - \epsilon/2$.
\end{lemma}
\begin{proof}
\newcommand{\maxvar}{\ensuremath{m}}
For any integer $\maxvar \geq 2$, a straightforward computation
yields that
\begin{equation*}
\Prob(\Qnum \leq \maxvar) = \sum_{t=1}^{\maxvar} P_a \, (1-P_a)^{t-1}
= 1 - (1-P_a)^\maxvar.
\end{equation*}
\noindent Therefore we have
\begin{eqnarray*}
\Prob(\max_k \Qnum_k \leq \maxvar) & = &  \big[1 - (1-P_a)^\maxvar \big]^K \\
& = & \big[1 - \exp (\maxvar \log(1-P_a)) \big]^K~.
\end{eqnarray*}

We want to choose $\maxvar = \maxvar(K, \epsilon)$ such that this
probability is greater than or equal to $1 - \epsilon/2$.  First
set \mbox{$\maxvar = -\rho \frac{\log K}{\log(1-P_a)}$,} where $\rho$ is
to be determined.  Then we have
\begin{equation*}
\Prob(\max_k \Qnum_k \leq \maxvar) = \big[1 - 1/K^\rho \big]^K~.
\end{equation*}
We now need to choose $\rho > 1$ such that
\begin{equation*}
\big[1 - 1/K^\rho \big]^K  \; \geq \; 1 - \epsilon/2~,
\end{equation*}
or equivalently, such that
\begin{eqnarray*}
1 - \big[1 - 1/K^\rho \big]^K & \leq & \epsilon/2~.
\end{eqnarray*}
Without loss of generality, let $K$ be even.  Then by convexity,
we have $(1-y)^K \geq 1 - Ky$.  Apply this with $y = 1/K^\rho$ to
obtain
\begin{eqnarray*}
1 - \big[1 - 1/K^\rho \big]^K & \leq & 1/K^{\rho-1}.
\end{eqnarray*}
Hence we need to choose $\rho \geq \log (2/\epsilon)/\log K + 1$
for the bound to hold.  Thus, if we set
\begin{eqnarray*}
\maxvar = - \rho \frac{\log K}{\log(1-P_a)} =
\order(\log(1/\epsilon) + \log K)
\end{eqnarray*}
then with probability greater than $1-\epsilon/2$, all $K$ rounds
of the protocol will use less than $\maxvar$ rounds of rejection.

\end{proof}

%%%%%%%%%%%%%%%%%%%%%%%
%%% GOSSIP ANALYSIS %%%
%%%%%%%%%%%%%%%%%%%%%%%
\subsection{Averaging with gossip}

As with averaging algorithms based on pairwise updates
\cite{BoydInfocom}, the convergence rate of our method is
controlled by the second largest eigenvalue $\eigtwo(\Wmat)$ of
the matrix
\begin{align*}
\Wmat & \defn I + \frac{1}{2 n} \left[\Pmat + \Pmat^T - D\right]~,
\end{align*}
where $D$ is diagonal with entries $D_i = (\sum_{j=1}^n[\Pmat_{ij}
+ \Pmat_{ji}])$.  The $(i,j)$-th entry of the matrix $\Pmat$ is
the probability that node $i$ exchanges values with node $j$.
Without rejection sampling, $P_{ij} = a_j$, and with rejection
sampling, $P_{ij} = q_j$.  With this notation, we are now equipped
to state and prove the main result of the paper:
\begin{theorem}
\label{Wbound} The geographic gossip protocol with rejection
threshold $\tau = \Theta(n^{-1}$)  has an averaging time
\begin{eqnarray}
T_{ave}(n, \epsilon) & = & \order \big(n \log (1/\epsilon) \big).
\end{eqnarray}
\end{theorem}
\noindent \begin{proof} To establish this bound, we exploit
Theorem 3 of \cite{BoydInfocom}, which states that the
$\epsilon$-averaging time is
\begin{equation}
T_{ave}(\epsilon, P) = \Theta \left( \frac{\log
\epsilon^{-1}}{\log \lambda_2(W)^{-1}} \right)~. \label{tave_bound}
\end{equation}
Thus, it suffices to prove that $\log \lambda_2(W) = \Omega(1/n)$
to establish the claim.

The probability of any sensor choosing sensor $v$ is just $q_v$,
so that the matrix $P = \ones \qvec^T$.   Note that the diagonal
matrix $D$ has entries
\begin{equation*}
D_i = \sum_{j=1}^n (P_{ij} + P_{ji}) = \sum_{j=1}^{n} q_j +
\sum_{j=1}^{n} q_i = 1 + n q_i~.
\end{equation*}
Thus, we can write $W$ in terms of outer products as:
\begin{equation}
\label{EqnWdecom} W = \left(I - \diag(\ones + n \qvec)\right) +
\frac{1}{2n} (\ones \qvec^T + \qvec \ones^T)~.
\end{equation}
Note that the matrix $W$ is symmetric and positive semidefinite.

We claim that the second largest eigenvalue $\lambda_2(W) =
\order(1 - c/n)$, for some constant $c$.  By Taylor series
expansion, this will imply that $\log \lambda_2(W) =
\Theta(n^{-1})$ as desired. To simplify matters, we transform the
problem to finding the maximum eigenvalue of an alternative
matrix. Since $W$ is doubly stochastic, its largest eigenvalue is
$1$ and corresponds to the eigenvector $v_1 = n^{-1/2} \ones$.
Consider the matrix $W' = W - \frac{1}{n^2} \ones \ones^T$; using
equation~\eqref{EqnWdecom}, it can be decomposed as
\begin{equation*}
W' = D' + Q',
\end{equation*}
where $D' = (I - (2n)^{-1} \diag(\ones + n \qvec))$ is diagonal
and
\[
Q' = \frac{1}{2n} \; (\ones(\qvec - n^{-1} \ones)^T + (\qvec -
n^{-1} \ones) \ones^T)
\]
is symmetric.

Note that by construction, the eigenvalues of $W'$ are simply
\[
\lambda(W) = \big \{(1 - \frac{1}{n}, \lambda_2(W), \ldots,
\lambda_n(W) \big \}.
\]
On one hand, suppose that $\lambda_1(W') > \lambda_2(W)$; in this
case, then $(1 - \frac{1}{n}) > \lambda_2(W)$ and we are done.
Otherwise, we have
\begin{equation*}
\lambda_1(W') = \lambda_2(W)~.
\end{equation*}
Note that $W'$ is the sum of a diagonal matrix and a symmetric
matrix with small entries.  Weyl's theorem
\cite[p.181]{HornJohnson} guarantees that
\begin{equation*}
\lambda_1(W') \le \lambda_1(D') + \lambda_1(Q') \le \left(1 -
\frac{1}{2n}\right) + \lambda_1(Q')~.
\end{equation*}
\noindent It is therefore sufficient to bound $\lambda_1(Q')$. We
do so using the Rayleigh-Ritz theorem \cite[p.176]{HornJohnson},
the Cauchy-Schwartz inequality, and Lemma~\ref{rejnorm} as
follows:
\begin{align*}
\lambda_1(Q') &= \max_{\yvec : \|\yvec\|_2 = 1} \yvec^T Q' \yvec \\
&= \frac{1}{2n} \max_{\yvec : \|\yvec\|_2 = 1} \yvec^T
(\ones(\qvec - n^{-1} \ones)^T + (\qvec - n^{-1} \ones) \ones^T
\yvec \\ &= \frac{1}{n} \max_{\yvec : \|\yvec\|_2 = 1} \yvec^T
\ones (\qvec - n^{-1} \ones)^T \yvec \\ &\le \frac{1}{n}
\max_{\yvec : \|\yvec\|_2 = 1} \|\yvec\|_2 \cdot \|\ones\|_2 \cdot
\|\qvec - n^{-1} \ones\|_2 \cdot \|\yvec\|_2 \\ &\le \frac{1}{n}
\left(1 \cdot \sqrt{n} \cdot \frac{1}{\sqrt{n}} \epsilon \right)
\\ &= \frac{1}{n} \epsilon
\end{align*}
\noindent  Now we have the total bound
\begin{equation}
\lambda_1(W') \le (1 - \frac{1}{2n}) + \frac{1}{n} \epsilon
\end{equation}
\noindent We can choose $\epsilon < 1/4$ using Lemma \ref{rejnorm}
to get the desired bound.
\end{proof}

The preceding theorem shows that by using rejection sampling we
can bound the convergence time of the gossip algorithm.  We can
therefore bound the number of radio transmissions required to
estimate the average:
\begin{corollary}
The expected number of radio transmissions required for our gossip
protocol on the geometric random graph $G(n, \sqrt{\frac{\log
n}{n}})$ is upper bounded as \beq \EE(n,\epsilon)= \order
\left(\frac{n^{3/2}}{\sqrt{\log n}} \log \epsilon^{-1} \right).
\eeq \noindent Moreover, with probability greater than $1 -
\epsilon/2$, the maximum number of radio transmissions is upper
bounded
\begin{equation}
\dur(n, \epsilon) =  \order \biggr (\EE(n,\epsilon) \big[ \log n +
\log \epsilon^{-1} \big] \biggr).
\end{equation}
\end{corollary}
{\bf Remark:} Note that for $\epsilon = n^{-a}$ for any $a > 0$,
our bounds are of the form $\EE(n, 1/n^a) = \order (n^{3/2}
\sqrt{\log n})$ and $\dur(n, \epsilon) = \order (n^{3/2} \log^{3/2} n)$.

\begin{proof}
\noindent We just have to put the pieces together. If we assume an
asynchronous protocol, the cost per transmission pair is given by
the product of $O(\sqrt{n/\log n})$ from routing, $\Exs[\Qnum]$
from rejection sampling, and the averaging time $T_{ave}$.  From
Lemma~\ref{exp_rej}, $\Exs[\Qnum] = O(1)$.  Using
equation~\eqref{tave_bound} and Theorem~\ref{Wbound}, we can bound
$\log \lambda_2(W)^{-1}$ by $(1 - \lambda_2(W)) = O(n^{-1})$.
Thus, the expected number of communications is
\begin{equation}
\order \left(\sqrt{\frac{n}{\log n}} \Exs[\Qnum] n \log
\epsilon^{-1} \right) = \order \left(\frac{n^{3/2}}{\sqrt{\log n}}
\log(\epsilon^{-1}) \right)~.
\end{equation}
\noindent To upper bound the maximum number of transmissions with
high probability, we note that Lemma~\ref{max_rej} guarantees that
\[
\max_{k=1, \ldots, T_{ave}} \Qnum_k = \order(\log T_{ave} +
\log \epsilon^{-1})
\]
with high probability.  Using Theorem~\ref{Wbound}, we can see that
$\order(\log T_{ave} + \log \epsilon^{-1}) = \order (\log n + \log
\epsilon^{-1})$. Consequently, with probability greater than $1 -
\epsilon/2$,
\begin{equation}
\dur(n, \epsilon) = \order \biggr (\EE(n,\epsilon) \big[ \log n +
\log \epsilon^{-1} \big] \biggr)~.
\end{equation}
\end{proof}

\section{Simulations}
%Simulations and comparision
\label{simulations}

Note that the averaging time is defined in
equation~\eqref{ave_time_definition} is a conservative measure,
obtained by selecting the worst case initial field $x(0)$ for each
algorithm.  Due to this conservative choice, an algorithm is
guaranteed to give (with high probability) an estimated average
that is $\epsilon$ close to the true average for any choices of
the underlying sensor observations.  As we have theoretically
demonstrated, our algorithm is provably superior to standard
gossiping schemes in terms of this metric.  In this section, we
evaluate our geographic gossip algorithm experimentally on
specific fields that are of practical interest.  We construct
three different fields and compare geographic gossip to the
standard gossip algorithm with uniform neighbor selection
probability.  Note that for random geometric graphs, standard
gossiping with uniform neighbor selection has the same scaling
behavior as with optimal neighbor selection
probabilities~\cite{BoydInfocom}, which ensures that the
comparison is fair.

\begin{figure}
\label{experiments}
\includegraphics[width=8cm]{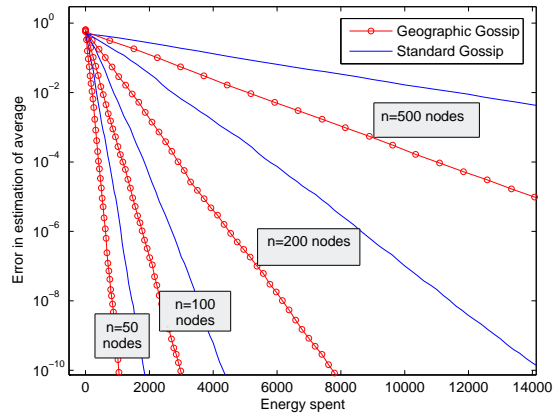}
\caption{Estimation accuracy versus total spent energy for a
linearly varying field.} \label{FigA}
\end{figure}

\begin{figure}
\includegraphics[width=8cm]{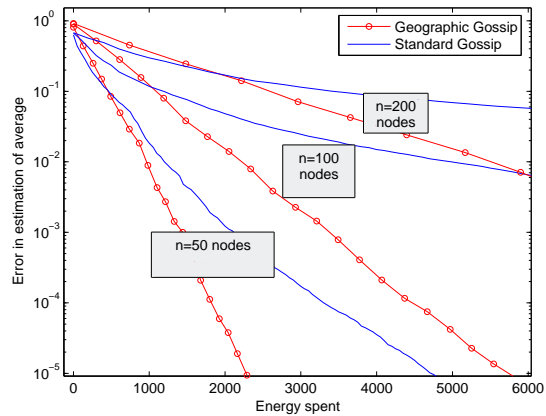}
\caption{Estimation accuracy versus total spent energy for a
smooth field modeling temperature.} \label{FigB}
\end{figure}

\begin{figure}
\includegraphics[width=8cm]{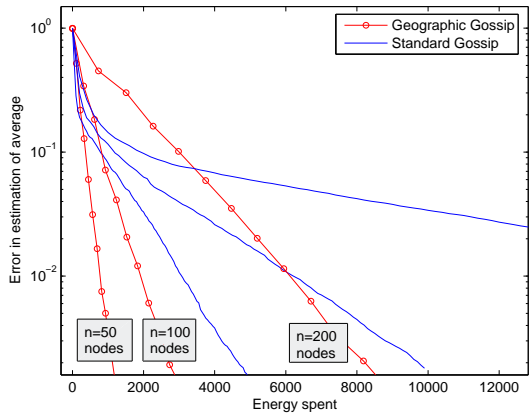}
\caption{Estimation accuracy versus total spent energy for a field
  which is zero everywhere except in one node.} \label{FigC}
\end{figure}

Figures~\ref{FigA} through~\ref{FigC} illustrates how the cost of
each algorithm behaves for various fields and network sizes. The
error in the average estimation is measured by the normalized
$\ell_2$ norm $\frac{\|x(k)- x_{ave} \ones \|}{\|x(0)\|}$. On the
other axis we plot the total number of radio transmissions
required to achieve the given accuracy.  Figure~\ref{FigA}
demonstrates how the estimation error behaves for a field that
varies linearly across one axis of the unit square.  In
Figure~\ref{FigB}, we use a field that is created by placing three
temperature sources in the unit square and smooth the field by a
simple process that models temperature diffusion.  Finally, in
Figure~\ref{FigC}, we use a field that is zero everywhere except
in one node.  For this field, the geographic gossip protocol
significantly outperforms the standard gossip protocol as the
network size and time increase, except for large estimation
tolerances ($\epsilon \approx 10^{-1}$) and few rounds.

As would be expected, simple gossip is capable of computing local
averages quite fast. Therefore, when the field is sufficiently
smooth, or when the averages in local node neighborhoods are close
to the global average, simple gossip might generate approximate
estimates which are closer to the true average with a smaller
number of transmissions. For these cases however, finding the
global average will not be useful in the first place.  In all our
simulations, the energy gains obtained by using geographic gossip
were significant and asymptotically increasing for larger network
sizes as our theoretical results suggest.

\section{Conclusions}
\label{SecDiscussion}

In this paper we have proposed a novel gossiping algorithm for
computing averages in networks in a completely distributed and
robust way. Geographic gossip computes the averages faster than
standard nearest neighbor gossip because it is using geographic
knowledge to quickly diffuse information everywhere in the
network. It is not hard to see that our algorithm is efficient for
grids (computes the $1/n^\alpha$ average in $O(n^{1.5} \log n)$
transmissions) and other topologies that realistically model
wireless networks. Even if geographic routing cannot be performed,
similar gossip algorithms can be used for any network that can
support some form of routing to random nodes. Essentially, we can
have nearest-neighbor gossip happening on the overlay network
supported by random routing.

The proposed algorithm can be used instead of nearest neighbor
gossip in all the schemes that use consensus based aggregation and
will greatly reduce the communication cost. For example
\cite{Kalman, Localization, Xiao} use similar ideas for
localization, Kalman filtering and sensor fusion. In these
schemes, geographic gossip can be used instead of standard
nearest-neighbor gossip to improve energy consumption.

\section*{Acknowledgments}
The work of Alexandros D. G. Dimakis was supported by NSF Grants
CCR-0219722 and CCR-0330514. The work of Anand D. Sarwate was
supported in part by the NSF Grant CCF-0347298.  The work of
Martin J. Wainwright was supported by an Intel Corporation Grant
and Alfred P. Sloan Foundation Fellowship.

\end{document}